\documentclass[aps,prr,twocolumn,nofootinbib,,floatfix,superscriptaddress,preprintnumbers]{revtex4-2}

\pdfoutput=1
\usepackage{amsmath,amssymb}
\usepackage{amsfonts}
\usepackage{bm,bbm}
\usepackage{graphicx}
\usepackage{mathrsfs}
\usepackage{slashed}
\usepackage{booktabs}
\usepackage{tabu}
\usepackage[dvipsnames]{xcolor}
\usepackage[normalem]{ulem}
\usepackage{tikz}
\usepackage{soul}
\usepackage[colorlinks=true,linkcolor=blue]{hyperref}
\usepackage{xcolor}
\usepackage{ulem}
\usepackage{array}
\usepackage{verbatim}
\usepackage{epsfig}
\usepackage{multirow}

\def\be{\begin{equation}}
\def\ee{\end{equation}}
\def\bea{\begin{eqnarray}}
\def\eea{\end{eqnarray}}

\begin{document}

\preprint{ADP-25-34/T1296}

\title{Relaxed constraints for dark matter with axial coupling to a dark photon}

\author{X. G.~Wang}
\email{xuan-gong.wang@adelaide.edu.au}
\affiliation{ARC Centre of Excellence for Dark Matter Particle Physics and CSSM, Department of Physics, University of Adelaide, Adelaide, South Australia 5005, Australia}

\author{A. W.~Thomas}
\email{anthony.thomas@adelaide.edu.au}
\affiliation{ARC Centre of Excellence for Dark Matter Particle Physics and CSSM, Department of Physics, University of Adelaide, Adelaide, South Australia 5005, Australia}
\date{\today}

\begin{abstract}
We present a model of the dark sector involving Dirac fermion dark matter, with axial coupling to a dark photon which provides a portal to Standard Model particles. In the non-relativistic limit, this implies that the dominant effective operator relevant to direct detection is ${\cal O}_8$. The resulting event rate for direct detection is suppressed by either the dark matter velocity or the momentum transfer. In this scenario there are much wider regions of the dark parameter space that are consistent with all of the existing constraints associated with thermal relic density, direct detection and collider searches. 
\end{abstract}

\date{\today}
\maketitle


\section{Introduction}
Although its existence has been confirmed in a multitude of ways~\cite{Cirelli:2024ssz}, the nature of dark matter (DM) remains a complete mystery. Over the past a few decades, an enormous effort has been devoted to searches for the weakly interacting massive particles (WIMPs), which appeared to be the most promising dark matter candidates, because of the so-called ``WIMP miracle"~\cite{Bertone:2004pz, Jungman:1995df, Goodman:1984dc}. The enthusiasm for WIMPs has cooled down as direct searches around the world have placed increasingly stringent constraints on the mass and couplings of such particles, focusing on the standard spin-independent (SI)~\cite{CRESST:2019jnq, CRESST:2019axx, DarkSide-50:2022qzh, XENON:2018voc, XENON:2020gfr, PandaX-4T:2021bab, LZ:2022lsv, LZ:2024zvo} and spin-dependent (SD)~\cite{PandaX-II:2016wea, LUX:2017ree, XENON:2019rxp,PICO:2019vsc, LZ:2024zvo} interactions.

 In addition to the constraints arising from direct searches, the observed relic density requires a sufficiently large cross section for dark matter annihilation to Standard Model (SM) particles ($s$-channel) to avoid overabundance. This can lead to tension with direct detection results because the corresponding DM-nucleus scattering cross sections ($t$-channel) may well exceed the upper limits set by direct detection. Combined constraints from the dark matter thermal relic density and direct detection have been employed to test various dark matter models~\cite{Zheng:2010js, Yu:2011by, Balan:2024cmq}.

Possible solutions to the challenge posed by this tension have been extensively investigated within the dark photon portal~\cite{Izaguirre:2015yja, Feng:2017drg, Krnjaic:2025noj, Alonso-Gonzalez:2025xqg, Wang:2025clh, Lee:2025lko}. The resonance regime is of particular interest in the case when $2 m_{\rm DM} < m_{A'}$, where $m_{\rm DM}$ and $m_{A'}$ are the dark matter and dark photon masses, respectively. The dark matter annihilation cross section is dominated by the $s$-channel resonance contribution, which is significantly enhanced near the region of $2 m_{\rm DM} \approx m_{A'}$, so that the dark couplings are not necessarily too large. However, most analyses have considered vector coupling of the dark photon to dark matter particles. In the non-relativistic limit, this results in a standard spin-independent (SI) interaction, for which there are very limited regions of the dark parameter space consistent with all existing constraints~\cite{Izaguirre:2015yja, Feng:2017drg, Krnjaic:2025noj, Alonso-Gonzalez:2025xqg, Wang:2025clh}. Indeed these constraints are so tight for dark Dirac fermions, that they are essentially ruled out. The situation is only slightly better for other scenarios such as complex scalar~\cite{Wang:2025clh}, pseudo-Dirac  and asymmetric dark matter. 
Alternatively, the secluded region of $m_{\rm DM} > m_{A'}$ is also promising for alleviating the tension between thermal relic and direct detection~\cite{Pospelov:2007mp}. In this case, the relic density is governed by $t$-channel annihilation, ${\rm DM} + {\rm DM} \to A' + A'$, while the direct detection cross sections could be suppressed.

Collider searches have also placed strong constraints on the dark photon parameters,  without~\cite{BaBar:2014zli, LHCb:2019vmc, CMS:2019buh} and with~\cite{BaBar:2017tiz, CMS:2021far, CMS:2024zqs} couplings to dark matter particles. In the latter case, the collider searches also set upper bounds on the standard SI and SD DM--nucleon cross sections, albeit strongly depending on the chosen coupling and model scenario~\cite{CMS:2021far, CMS:2024zqs}.

In this paper, we investigate a model of the dark sector, comprising a dark photon, which is taken to be a U(1) gauge boson kinematically mixed with the $B$ of the Standard Model and which has axial-vector coupling to Dirac fermion dark matter with mass up to 1 TeV. In this case, the dominant effective operator for dark matter nucleus scattering in the non-relativistic limit is ${\cal O}_8$, which depends on the velocity of the dark matter particle and the momentum transfer~\cite{Anand:2013yka, Fitzpatrick:2012ix}. The resulting DM--nucleon cross sections are highly suppressed and, as a result, are expected to satisfy the strong constraints from direct detection~\cite{Dent:2015zpa, DEramo:2016gos}. However, the $s$-channel dark matter annihilation cross section will be also suppressed, therefore requiring larger values of the dark couplings to avoid overabundance. Therefore, it is non-trivial to explore the regions of dark parameter space that are consistent with the constraints of both direct and indirect detection and the observed relic density. As demonstrated below, the resonance regime remains viable, yielding allowed regions considerably broader than those identified in previous models.

We begin with a brief review of the dark photon formalism in Sec.~\ref{sec:DP}. The existing constraints on dark matter from thermal relic density, direct detection, and collider searches are given in Sec.~\ref{sec:constraints}. We present the allowed regions of the dark parameter space that are consistent with these constrains in Sec.~\ref{sec:results}. Finally, we summarize our conclusions in Sec.~\ref{sec:conculsion}.

\section{Dark photon formalism}
\label{sec:DP}
The dark photon, $A'$, is usually introduced as an extra $U(1)$ gauge boson, interacting with SM particles through kinetic mixing with the hypercharge $B$ boson~\cite{Fayet:1980ad, Fayet:1980rr, Holdom:1985ag, Okun:1982xi}. Here, we introduce axial-vector coupling of the dark photon to Dirac fermion dark matter $\chi$~\footnote{Our phenomenological model may be derived from a UV complete theory~\cite{Roy:2025inq}, but the details of that do not affect the phenomenological analysis. A minimal completion involves an additional fermion species ($\psi$) and a complex singlet scalar $S$. By imposing opposite $U'(1)$ charges, $Q'(\chi_L) = +q,\ Q'(\chi_R) = -q$, and $Q'(\psi_L) = -q,\ Q'(\psi_R) = +q$, one ensures an anomaly-free U'(1) with the fermions and the resulting interaction is purely axial. After $S$ acquires a nonzero vacuum expectation value (VEV), the dark photon and dark fermions obtain masses. The lighter dark fermion is stable and therefore the dark matter candidate. Our results presented in this work can be applied in the case of a small mixing between $\chi$ and $\psi$.}
\bea
\label{eq:L_chi}
{\cal L} &=& - \frac{1}{4} F'_{\mu\nu} F'^{\mu\nu} + \frac{1}{2} m^2_{A'} A'_{\mu} A'^{\mu} 
+ \frac{\epsilon}{2 \cos\theta_W} F'_{\mu\nu} B^{\mu\nu} \nonumber\\
&& + \bar{\chi} ( i \slashed{\partial} - m_{\chi} ) \chi +  g_{\chi} \bar{\chi} \gamma^{\mu} \gamma_5 \chi A'_{\mu}\, ,
\eea
where $\epsilon$ is the mixing parameter and $\theta_W$ is the Weinberg angle. 

The mixing term can be removed through field redefinitions and the physical $Z$ and dark photon $A_D$ can be written in terms of the unmixed fields $\bar{Z}$ and $A'$,
\be
\begin{pmatrix}
Z_{\mu}\\
A_{D\mu}\\
\end{pmatrix} 
=
\begin{pmatrix}
\cos\alpha & \sin\alpha \\
- \sin\alpha & \cos\alpha\\
\end{pmatrix}
\begin{pmatrix}
\bar{Z}_{\mu}\\
A'_{\mu}\\
\end{pmatrix} \, ,
\ee
with masses being~\cite{Kribs:2020vyk}
\begin{eqnarray}
\label{eq:m_Z_AD}
M^2_{Z, A_D} &=& \frac{m_{\bar{Z}}^2}{2} [ 1 + \epsilon_W^2 + \rho^2 \nonumber\\
&& \pm {\rm sign}(1-\rho^2) \sqrt{(1 + \epsilon_W^2 + \rho^2)^2 - 4 \rho^2} ] \, .
\end{eqnarray}
Here, $\alpha$ is the $\bar{Z}-A'$ mixing angle,
\begin{eqnarray}
\tan \alpha &=& \frac{1}{2\epsilon_W} \Big[ 1 - \epsilon^2_W - \rho^2 \nonumber\\
&& - {\rm sign}(1-\rho^2) \sqrt{4\epsilon_W^2 + (1 - \epsilon_W^2 - \rho^2)^2} \Big] \, , 
\end{eqnarray}
with
\begin{eqnarray}
\epsilon_W &=& \frac{\epsilon \tan{\theta_W}}{\sqrt{1 - \epsilon^2 / \cos^2\theta_W}}\, ,\nonumber\\
\rho &=& \frac{m_{A'}/m_{\bar{Z}}} {\sqrt{1 - \epsilon^2 / \cos^2\theta_W}}\, .
\end{eqnarray}
The mass difference is always finite for non-zero $\epsilon$, $|M^2_Z - M^2_{A_D}|\ge 2 |\epsilon| m^2_{\bar Z}$, yielding the so-called ``eigenmass repulsion" region in which the dark photon parameters are not accessible~\cite{Kribs:2020vyk}.

The couplings of the physical dark photon,  $A_D$, to SM fermions (in units of $e = \sqrt{4\pi\alpha_{\rm em}}$) are given by~\cite{Kribs:2020vyk}
\begin{eqnarray}
\label{eq:C_AD}
C_{A_D}^v &=& - (\sin\alpha + \epsilon_W \cos\alpha) C_{\bar{Z}}^v + \epsilon_W \cos\alpha \cot \theta_W C_{\gamma}^v ,\nonumber\\
C_{A_D}^a &=& - (\sin\alpha + \epsilon_W \cos\alpha) C_{\bar{Z}}^a 
\, .
\end{eqnarray}
The Standard Model couplings of the $Z$ boson, $C^v_{\bar Z}$ and $C^a_{\bar Z}$, will be shifted to the physical ones,
\bea
\label{eq:C_Z}
C_{Z}^v &=& (\cos\alpha - \epsilon_W \sin\alpha) C_{\bar{Z}}^v + 
\epsilon_W \sin\alpha \cot \theta_W C_{\gamma}^v ,\nonumber\\
C_{Z}^a &=& (\cos\alpha - \epsilon_W \sin\alpha) C_{\bar{Z}}^a\, ,
\eea 
where $C^v_{\gamma} = Q_f$ is the electromagnetic coupling, and
\be
C^v_{\bar Z} \sin 2\theta_W = T^f_3 - 2 Q_f \sin^2\theta_W\, ,\ C^a_{\bar Z} \sin 2\theta_W = T^f_3\, ,
\ee
with  $\{ T^e_3, T^u_3, T^d_3\} = \{ -1/2, 1/2, -1/2\}$ and $\{ Q_e, Q_u, Q_d\} = \{ -1, 2/3, -1/3\}$.

Both the dark photon $A_D$ and the $Z$ boson couple to dark matter particles $\chi$,
\bea
\label{eq:C-DM}
C^a_{A_D,\chi} &=& \frac{g_{\chi} \cos\alpha}{\sqrt{1 - \epsilon^2/\cos\theta^2_W}}\, ,\nonumber\\
C^a_{Z,\chi} &=& \frac{g_{\chi} \sin\alpha}{\sqrt{1 - \epsilon^2/\cos\theta^2_W}}\, .
\eea
The decay width of the dark photon is
\be
\Gamma_{A_D} = \Gamma_{A_D \to {\rm SM}} + \Gamma_{A_D \to \bar{\chi}\chi}\, ,
\ee
where
\bea
\label{eq:Gamma-AD}
\Gamma_{A_D \to {\rm SM}} &=&  \sum_{f} N_C^f \cdot \frac{M_{A_D} \alpha_{\rm em}}{3} \bigg\{\left(1 + \frac{2m_{f}^2}{M_{A_D}^2}\right) (C^v_{A_D,f})^2 \nonumber\\
&& + \left(1 - \frac{4m_{f}^2}{M_{A_D}^2}\right) (C^a_{A_D,f})^2\bigg\}\sqrt{1 - \frac{4m_{f}^2}{M_{A_D}^2}}\, ,\nonumber\\
\Gamma_{A_D \to \bar{\chi}\chi} &=&\frac{M_{A_D} (C^a_{A_D,\chi})^2}{12\pi} \left(1 - \frac{4 m_{\chi}^2}{M_{A_D}^2}\right) \sqrt{1 - \frac{4m_{\chi}^2}{M_{A_D}^2}}\, ,\nonumber\\
\eea
with $N_C^f = 1$ for leptons and $N_C^f = 3$ for quarks. 

It is crucial that the parameters of the model are consistent with the set of precise measurements of a select set of observables associated with the $Z$-boson, known as the electro-weak precision observables. The original analysis of Curtin {\em et al.,}~\cite{Curtin:2014cca} was recently updated by Loizos {\em et al.}~\cite{Loizos:2023xbj} and we will show those constraints explicitly.

\section{Constraints on dark matter}
\label{sec:constraints}
In this section we discuss the existing constraints on dark matter from thermal relic density, indirect and direct detection, and collider searches.

\subsection{Thermal relic density}
In the framework of thermal freeze out, the evolution of the dark matter number density, $n_{\rm DM}$, is governed by~\cite{Cirelli:2024ssz}
\be
\dot{n}_{\rm DM} + 3 H n_{\rm DM} = \langle \sigma v \rangle [(n^{\rm eq}_{\rm DM})^2 - n^2_{\rm DM}]\, ,
\ee
where $H$ is the Hubble constant and $n^{\rm eq}_{\rm DM}$ is the number density that DM particles would have in thermal equilibrium. $\langle \sigma v \rangle$ is the thermally-averaged cross section of dark matter annihilation to SM particles~\cite{Gondolo:1990dk, Griest:1990kh}. 

In the limit $\epsilon \ll 1$, the physical couplings $C^v_{A_D,f} \sim Q_f \epsilon$ and $C^a_{A_D,\chi} \sim g_{\chi}$. Therefore, the dark matter annihilation cross section depends on the dimensionless variable~\cite{Izaguirre:2015yja, Filippi:2020kii}
\be
\label{eq:y}
y = \epsilon^2 \alpha_D \left( \frac{m_{\chi}}{M_{A_D}} \right)^4\, ,
\ee
where $\alpha_D = g^2_{\chi}/4\pi$. The observed relic density~\cite{ParticleDataGroup:2024cfk}, 
\be
\label{eq:Omega}
\Omega = \Omega_{\rm DM} h^2 = 0.1200 \pm 0.0012\, ,
\ee
is usually applied to set lower limits on $y$ to avoid overabundance. 

A general feature is that the required values of $y$ will decrease as the dark photon mass approaches the dark matter threshold, with the mass ratio $R = M_{A_D} / m_{\chi} \approx 2$~\cite{Izaguirre:2015yja, Feng:2017drg, Krnjaic:2025noj, Alonso-Gonzalez:2025xqg, Wang:2025clh}. This is often referred to as the resonance region. In addition, the resonance contribution associated with the $Z$ boson will also be significant; that is, when $2 m_{\chi} \approx M_Z$~\cite{Wang:2025clh}.


\subsection{Dark matter indirect detection}
Dark matter indirect detection, such as the Cosmic Microwave Background (CMB)~\cite{Planck:2018vyg}, Fermi-LAT~\cite{Fermi-LAT:2015att} and AMS-02~\cite{AMS:2019rhg}, could place stringent constraints on the annihilation rate~\cite{Dutta:2022wdi}. 

In the case of $s$-wave annihilation, the cross section required by the relic density is~\cite{Steigman:2012nb}
\be
\langle \sigma v \rangle \approx 3 \times 10^{-26} {\rm cm}^3 s^{-1}\, ,
\ee
which exceeds the upper bounds of indirect detection for dark matter mass up to a few hundred GeV~\cite{Dutta:2022wdi}.

However, for the axial-vector interaction considered here, the dark matter annihilation is $p$-wave. Therefore, $\langle \sigma v \rangle$ in the present day galaxies is suppressed by $v^2 \sim {\cal O}(10^{-6})$, which clearly satisfies those indirect constraints.

\subsection{Dark matter direct detection}
Given that the momentum transfer, $q^2$, is small, we can derive the leading effective four-fermion interaction by integrating out the heavy degrees of freedom
\be
\label{eq:L-EFT}
{\cal L}_{\rm EFT} = C^N_{AV} \bar{\chi} \gamma^{\mu} \gamma_5 \chi \bar{N} \gamma_{\mu} N + C^N_{AA} \bar{\chi} \gamma^{\mu} \gamma_5 \chi \bar{N} \gamma_{\mu} \gamma_5 N\, ,
\ee
where 
\bea
\label{eq:CAV-CAA}
C^N_{AV} &=& \frac{C^a_{A_D,\chi}\cdot e C^v_{A_D,N}}{M^2_{A_D}} + \frac{C^a_{Z,\chi} \cdot e C^v_{Z,N}}{M^2_Z}\, ,\nonumber\\
C^N_{AA} &=& - \frac{C^a_{A_D,\chi} \cdot e C^a_{A_D,N}}{M^2_{A_D}} - \frac{C^a_{Z,\chi} \cdot e C^a_{Z,N}}{M^2_Z}\, .
\eea
The couplings to the nucleon, $C_{A_D,N}^{v,a}$ and $C_{Z,N}^{v,a}$, can be derived as linear combinations of the couplings to the $u$ and $d$ quarks,
\bea
C_{A_D(Z),p}^{v,a} &=& 2 C_{A_D(Z),u}^{v,a} + C_{A_D(Z),d}^{v,a} \, ,\nonumber\\
C_{A_D(Z),n}^{v,a} &=& C_{A_D(Z),u}^{v,a} + 2 C_{A_D(Z),d}^{v,a}\, .
\eea
The non-relativistic reduction of Eq.~(\ref{eq:L-EFT}) leads to 
\be
{\cal L}_{\rm NREFT} = \sum_{N=p,n} \left( c^N_8 {\cal O}_8 + c^N_9 {\cal O}_9 + c^N_4 {\cal O}_4 \right)\, ,
\ee
where the effective operators are~\cite{Anand:2013yka} 
\bea
\label{eq:O8-O9-O4}
{\cal O}_8 &=& \vec{S}_{\chi} \cdot \left(\vec{v} \, + \, \frac{\vec{q}}{2 \mu_{\chi N}} \right) \, ,\nonumber\\
{\cal O}_9 &=& i \vec{S}_{\chi} \cdot \left( \vec{S}_{N} \times \frac{\vec{q}}{m_N} \right)\, ,\nonumber\\
{\cal O}_4 &=& \vec{S}_{\chi} \cdot \vec{S}_N\, ,
\eea
and the couplings (with dimension $M^{-2}$) are
\be
\label{eq:cN}
c^N_8 = c^N_9 = 2 C^N_{AV}\, ,\ \ \ c^N_4 = - 4 C^N_{AA}\, .
\ee
In the region of dark parameter space of interest, the couplings $c^N_4$, responsible for spin-dependent scattering,  are negligibly small because of an extremely strong cancellation, independent of $M_{A_D}$, between the dark photon and the $Z$ boson terms in $C^N_{AA}$ in Eq.~(\ref{eq:CAV-CAA}) (see Appendix~\ref{sec:O4}). As shown in Appendix~\ref{sec:O8-O9}, the operator ${\cal O}_8$ dominates the event rate of direct detection.

While most direct searches have focused on providing limits on the dark matter proton cross section based upon the standard SI and SD interactions, the Xenon Collaboration has placed exclusion limits on isoscalar dimensionless couplings for all elastic scattering EFT operators~\cite{XENON:2017fdd}. In the case of ${\cal O}_8$, it is assumed that $C^0_8 = C^p_8 = C^n_8$, which implies that, as a good approximation, the event rate $R \propto A^2 \cdot (C_8^0)^2$. However, in the dark photon model considered here, we find that $c^n_8 \ll c^p_8$. Therefore, we simply rescale the Xenon100 limits by a factor of $A^2/Z^2 = 131^2/54^2$. A precise determination should take into account the proportion of the seven most abundant isotopes of Xenon and the isospin asymmetric nuclear response functions. However, as we will see when the predictions of the present model are compared with the Xenon100 (and LZ) limits, the size of those corrections is far below the accuracy relevant at the present time.

The LZ Collaboration~\cite{LZ:2024zvo} has placed the most stringent constraints on the standard spin-independent (SI) cross sections. These are stronger than those of the Xenon-1T Collaboration~\cite{XENON:2020gfr} by approximately one order of magnitude. We make an estimate of the LZ limits on $c_8^p$ by assuming
\be
\label{eq:LZ}
\frac{(c_8^p)^2|_{\rm LZ}}{(c_8^p)^2|_{\rm Xenon100}} = \frac{\sigma_p^{\rm SI}|_{\rm LZ}}{\sigma_p^{\rm SI}|_{\rm Xenon1T}}\, .
\ee
%

\subsection{Constraints from collider searches}
As mentioned earlier, because of the $A'-Z$ mixing, strong indirect  constraints are provided by  electroweak precision observables~\cite{Hook:2010tw, Curtin:2014cca, Loizos:2023xbj},  measured at lepton (LEP, SLC) and hadron (Tevatron, LHC) colliders~\cite{ParticleDataGroup:2024cfk}. These constraints, which will be reported in the results section, typically lead to an upper limit on $\epsilon\sim {\cal O}(10^{-2})$. 

One can also derive exclusion limits on the parameter $\epsilon$ from an analysis of electron--proton deep inelastic scattering~\cite{Kribs:2020vyk, Thomas:2021lub, Yan:2022npz}. These are compatible with the EWPO bound for $M_{A_D} < 10\ {\rm GeV}$, while becoming weaker as the dark photon mass increases; even rising above 0.1 when $M_{A_D} > M_Z$. The one exception to these exclusion limits is the analysis of DIS data by Hunt-Smith {\em et al.}~\cite{Hunt-Smith:2023sdz}, who reported a strong indirect signal for a dark photon in the 2-6 GeV region.

The strongest limits on the mixing parameter $\epsilon$ come from $e^+ e^-$~\cite{BaBar:2014zli} and hadron colliders~\cite{LHCb:2019vmc, CMS:2019buh}, which typically assume that the dark photon only decays to SM final states. Recent analyses showed that these constraints could be significantly relaxed in light of potential couplings of the dark photon to dark matter ~\cite{Abdullahi:2023tyk, Felix:2025afw, Alonso-Gonzalez:2025xqg}. Taking the CMS constraints~\cite{CMS:2019buh} as an example, in the case of vector coupling the relaxed limits can be derived either from an analysis of cross sections~\cite{Felix:2025afw} or from a simple re-scaling by decay widths~\cite{Alonso-Gonzalez:2025xqg}. In the present work, we adopt the latter method to estimate the modified upper bounds
\be
\label{eq:relaxed-CMS}
\epsilon < \epsilon^{\rm CMS} \times \sqrt{ \frac{\Gamma_{A_D \to {\rm SM}} + \Gamma_{A_D \to \bar{\chi}\chi}}{\Gamma_{A_D\to {\rm SM}}} }\, ,
\ee
where the decay widths are given in Eq.~(\ref{eq:Gamma-AD}).

Constraints on $\epsilon$ from future experiment at Belle-II are expected to be much stronger for dark photon mass below a few GeV~\cite{Gori:2022vri}. The proposed experimental facilities, such as the FCC-ee~\cite{FCC:2018evy}, are expected to measure some of the electroweak observables with significantly increased precision, which could improve the current constraints on the dark sector in the heavy mass region~\cite{Wang:2025rsg}.

\section{Results}
\label{sec:results}
Earlier studies~\cite{Izaguirre:2015yja, Feng:2017drg, Krnjaic:2025noj} typically adopted a mass ratio of $R=3$ and then reduce it to $R\approx 2$ (resonance region). The dark coupling, $\alpha_D$, spans a wide range from the perturbativity limit of $1/2$ down to $\alpha_{\rm em}$ or smaller. To ensure our analysis thoroughly explores the most relevant parameter space, we select $R=3,\ 2.3,\ 2.05$ and $\alpha_D = 0.5,\ 0.05,\ 0.005$, as  in Ref.~\cite{Wang:2025clh}.

\begin{figure*}[!h]
\begin{center}
\includegraphics[width=\textwidth]{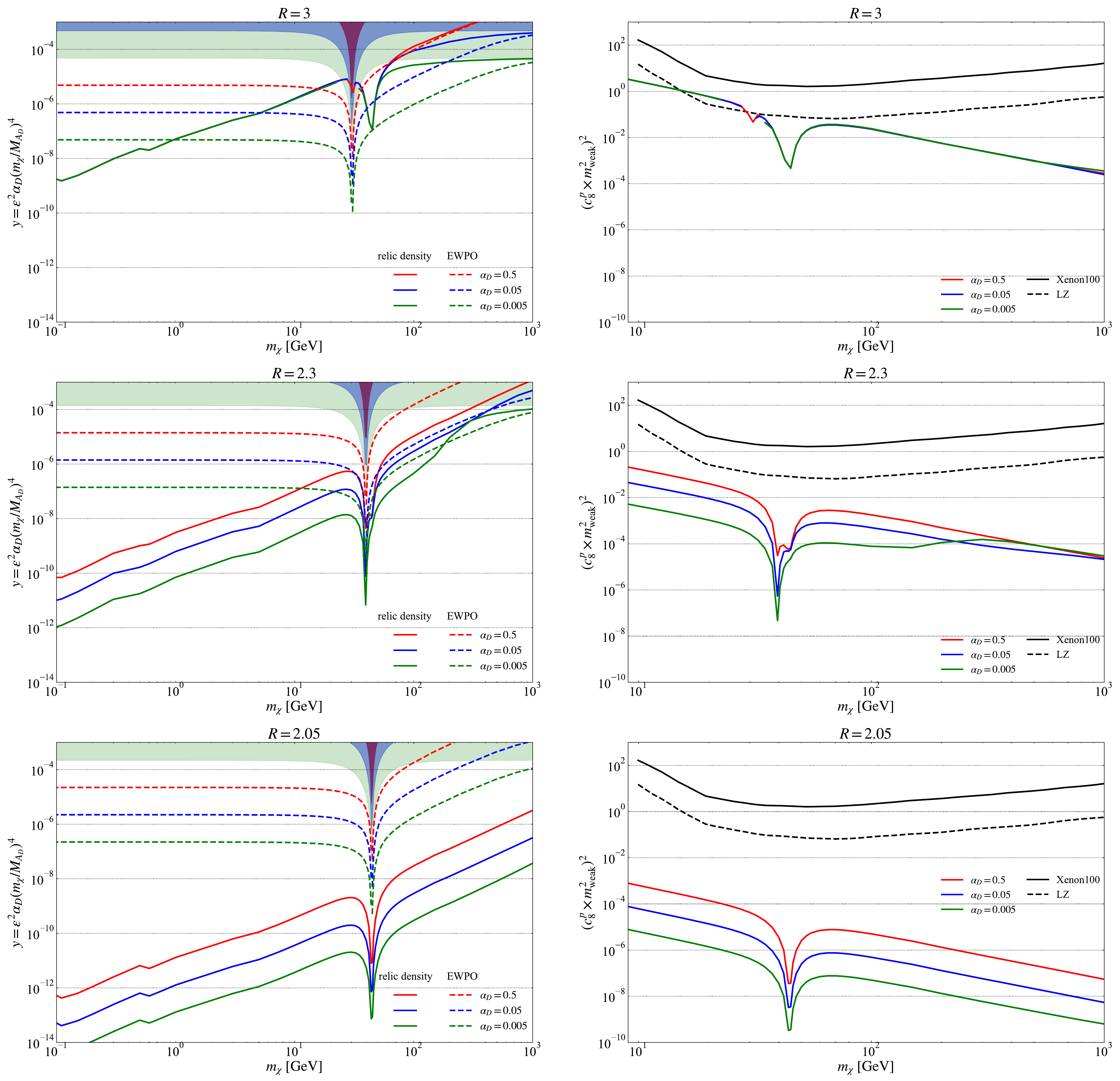}
\vspace*{-0.2cm}
\caption{(Left panels): The lower limits on $y$ from the thermal relic density (solid lines). The EWPO constraints (dashed lines) are derived by converting the exclusion limits on $\epsilon$ from Ref.~\cite{Curtin:2014cca, Loizos:2023xbj},  with $M_{A_D}$ being extended up to 3 TeV. The shaded areas are the eigen-mass repulsion regions~\cite{Kribs:2020vyk} corresponding to different values of $\alpha_D$, in which the dark photon parameters are not accessible.  (Right panels): the corresponding constraints on the dimensionless coupling of ${\cal O}_8$. The Xenon100 limits~\cite{XENON:2017fdd} have been relaxed by a factor of $A^2/Z^2$. The upper bounds from LZ are derived according to the rescaling in Eq.~(\ref{eq:LZ}).}
\label{fig:Dirac-axial}
\end{center}
\end{figure*}

For given values of $\alpha_D$ and the mass ratio $R$, we adjust the mixing parameter $\epsilon$ to generate the observed relic density in Eq.~(\ref{eq:Omega})~\cite{Christensen:2008py, Alloul:2013bka,Alguero:2023zol}. This sets the lower bounds on the variable $y$ required to avoid over-abundance, which are shown in Fig.~\ref{fig:Dirac-axial} (left panels). The exclusion constraints on $\epsilon$ from electroweak precision observables (EWPO), which are also shown there, are taken from Ref.~\cite{Curtin:2014cca, Loizos:2023xbj} and converted to the $y - m_{\chi}$ plane.

We then derive the corresponding couplings $c^p_8$ in Eq.~(\ref{eq:cN}) using the dark parameters determined from the thermal relic density,  which are converted to dimensionless couplings using $m^2_{\rm weak} = (246.2\ {\rm GeV})^2$, and compared with the upper limits from the Xenon100 Collaboration and the LZ Collaboration as shown in Fig.~\ref{fig:Dirac-axial} (right panels).

The most appealing feature of our model is that, even if the dark photon mass is far away from the resonant region with $R=3$, as shown in the right-hand panel of Fig.~\ref{fig:Dirac-axial}, the lower bounds on the coupling $c^p_8$ from the thermal relic density are below the upper limits set by direct detection, over a wide range of $m_\chi$. This is because the  direct detection event rate is suppressed by $v^2$ or $q^2$, where $v$ and $q$ are the dark matter velocity and the momentum transfer, respectively. For lower values of $R$, the lower bound from relic density considerations lies below the direct detection upper bound on $c^p_8$ for all $m_\chi$.

On the other hand, for $R=3$ the lower limits on $\epsilon$, or equivalently the variable $y$, exceed the upper bounds from electroweak precision observables over a broad range of $m_{\chi}$, in the GeV--TeV region. For $\alpha_D = 0.05$ and $0.005$, there are no solutions for $y$ because of the eigenmass repulsion when $m_{\chi}$ lies in the range $[28.5, 32.0]$ GeV and $[24.2, 34.9]$ GeV, respectively. Note that for $m_{\chi} < 60\ {\rm GeV}$, the lower limits of $y$ with $\alpha_D = 0.5,\ 0.05$ and $0.005$ coincide. In the high-mass region, there are small differences among these three cases because the values of $\epsilon$ are large and the small $\epsilon$ expansions of the physical couplings in Eqs.~(\ref{eq:C_AD}) and~(\ref{eq:C-DM}) are not valid.

As one moves a little closer to the dark matter threshold, with $R=2.3$, the relic density constraint allows smaller values of $y$ because of the enhanced contribution to the dark matter annihilation cross section. This allows one to escape the EWPO constraints for $\alpha_D = 0.5,\ 0.05$ and $0.005$, for a relatively wide range of dark photon masses. In addition, the resulting couplings $c^p_8$ lie well below the upper limits derived from direct detection.

In the resonance regime, with $R = 2.05$, the lower bounds on $y$ and the corresponding coupling $c^p_8$ decrease even further, leaving a much broader region of the dark parameter space that is consistent with the dark matter relic density and direct detection.

\begin{figure}[!h]
\begin{center}
\includegraphics[width=\columnwidth]{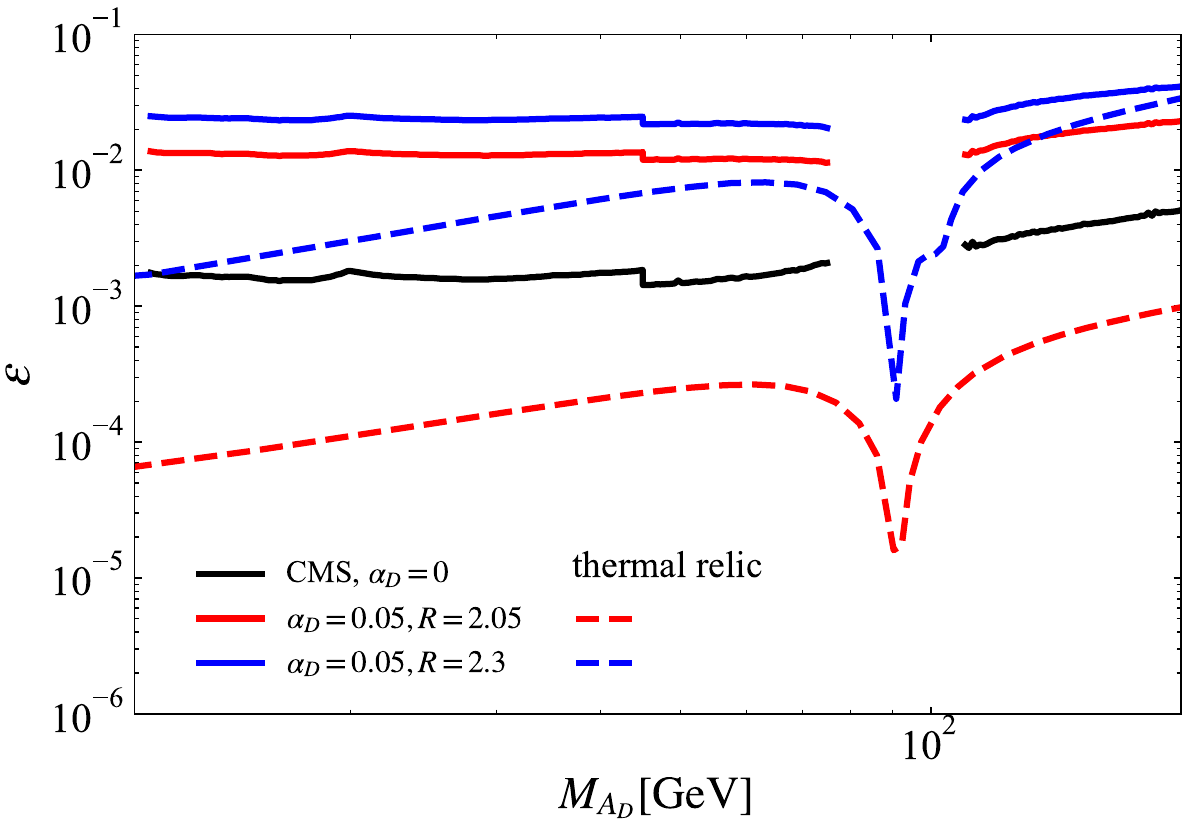}
\vspace*{-0.2cm}
\caption{The lower bounds on $\epsilon$ from the thermal relic density (dashed line), which are compared with the modified upper limits (solid lines) from the CMS collaboration~\cite{CMS:2019buh}, according to Eq.~(\ref{eq:relaxed-CMS}).}
\label{fig:eps-CMS}
\end{center}
\end{figure}

In the case where $\alpha_D \gg \alpha_{\rm em} \epsilon^2$, the dark photon decay width to dark matter will be much larger than that to SM particles. This has the effect of increasing the deduced limits on $\epsilon$ obtained by assuming that it decays only to SM particles. In Fig.~\ref{fig:eps-CMS} we show that the lower limits on $\epsilon$ obtained from the thermal relic density by taking $\alpha_D = 0.05$ for $R=2.3$ and $2.05$,  are consistent with the relaxed CMS constraints according to Eq.~(\ref{eq:relaxed-CMS}).
%

\section{Conclusion}
\label{sec:conculsion}
We have investigated a dark matter model consisting of a dark photon which interacts with a dark Dirac fermion through axial-vector coupling. Both the dark photon and the $Z$ boson contribute to the dark matter annihilation and dark matter--nucleon scattering processes. The leading effective operator in a non-relativistic reduction is ${\cal O}_8$. 
We demonstrated that the resonance regime is still viable with broad regions of dark parameter space that are consistent with all the constraints from dark matter relic density, direct detection and collider searches.

 In order to satisfy the EWPO constraints, it is still required that the ratio $R$ should not be too large. In all three cases with $R=3,\ 2.3$ and $2.05$, the event rate for DM--nucleus scattering naturally eludes the direct detection constraints because of suppression from the dark matter velocity and the momentum transfer. In the last two cases, the lower bounds on the mixing parameter $\epsilon$ are also consistent with the modified CMS limits.  

 Our findings suggest that WIMP dark matter is still a promising dark matter candidate, but with non-standard SI and SD interactions between the dark and ordinary matter.
 
\section*{Acknowledgments}
We would like to thank Raymond Volkas for a helpful discussion about UV completions. This work was supported by the University of Adelaide and the Australian Research Council through the Centre of Excellence for Dark Matter Particle Physics (CE200100008).


\appendix

\section{Cancellation in the evaluation  of ${\cal O}_4$}
\label{sec:O4}
In this section, we demonstrate the strong cancellation in the coupling $c^N_4$.

Substituting Eqs.~(\ref{eq:C_AD}-\ref{eq:C_Z}) and (\ref{eq:C-DM}) into Eq.~(\ref{eq:CAV-CAA}), the coupling to the proton is
\bea
\label{eq:CAA}
C^p_{AA} &=& \frac{g_{\chi} e (2 C^a_{\bar{Z}, u} + C^a_{\bar{Z},d})}{\sqrt{1 - \epsilon^2/\cos^2\theta_W}} 
\cdot 
\left\{ \frac{\cos\alpha (\sin\alpha + \epsilon_W \cos\alpha)}{M^2_{A_D}} \right.\nonumber\\
&& \left. - \frac{\sin\alpha (\cos\alpha - \epsilon_W \sin\alpha)}{M^2_{Z}} \right\}\, .
\eea
In the limit $\epsilon \ll 1$, we can get~\cite{Alonso-Gonzalez:2025xqg}~\footnote{The mixing parameter $\epsilon$ in Ref.~\cite{Alonso-Gonzalez:2025xqg} is equivalent to $- \epsilon/\cos\theta_W$ in our work.}
\bea
\sin\alpha &=& - \eta \frac{\epsilon}{\cos\theta_W} + {\cal O}(\epsilon^3)\, ,\nonumber\\
\cos\alpha &=& 1 - \frac{\eta^2}{2} \frac{\epsilon^2}{\cos\theta^2_W}\, ,
\eea
where $\eta = \sin\theta_W/(1 -r)$ with $r = M^2_{A_D}/M^2_Z$. Therefore, the terms in the bracket of Eq.~(\ref{eq:CAA}) become
\bea
&&\frac{1}{M^2_{A_D}} \cdot \left\{ ( \sin\alpha + \epsilon_W ) - \sin\alpha \cdot r \right\} \nonumber\\
&=& \frac{1}{M^2_{A_D}} \cdot \{ - \epsilon \tan\theta_W + \epsilon_W \}\, ,
\eea
which exactly cancel, leading to a vanishing coupling of $c^p_4$ at this order. Moreover, the cancellation also occurs for the coupling to the neutron, $c^n_4$.

\section{Contributions of ${\cal O}_8$ and ${\cal O}_9$}
\label{sec:O8-O9}
The event rate (per unit time) in direct detection can be written as
\be
\label{eq:dR-dER}
\frac{d R}{d E_R} = N_T \frac{\rho_{\chi}}{m_{\chi}} \cdot \langle \frac{d \sigma_{\chi T}}{d E_R} \rangle\, ,
\ee
where $\langle \frac{d \sigma_{\chi T}}{d E_R} \rangle$ denotes an average over the dark matter velocity distribution in the lab frame,
\be
\langle \frac{d \sigma_{\chi T}}{d E_R} \rangle\ = \int \frac{d \sigma_{\chi T}}{d E_R} \cdot v f(\vec{v},t) d^3 v\, . 
\ee
The differential DM-nucleus scattering cross section is~\cite{Anand:2013yka}
\be
\frac{d\sigma_{\chi T}}{d E_R} = \frac{2 m_T}{4 \pi v^2} \cdot \sum_{ij} \sum_{N,N'=p,n} c_i^{N} c_j^{N'} F_{ij}^{N,N'}(v^2,q^2)\, ,
\ee
where the nuclear response functions associated with the effective operators ${\cal O}_8$ and ${\cal O}_9$ are~\cite{Fitzpatrick:2012ix}~\footnote{The definition of ${\cal O}_9$ in Eq.~(\ref{eq:O8-O9-O4}) is different from that in Ref.~\cite{Fitzpatrick:2012ix} by a factor of $1/m_N$. Therefore, $F^{N,N'}_{9,9}$ and $F^{N,N'}_{8,9}$ are different from those in Ref.~\cite{Fitzpatrick:2012ix} by $1/m^2_N$ and $1/m_N$, respectively.}
\bea
\label{eq:F-NN}
F^{N,N'}_{8,8} &=& C(j_{\chi}) \frac{1}{4} \left( (v^2 - \frac{q^2}{4 m^2_T} ) F^{N,N'}_M + \frac{q^2}{m^2_N} F^{N,N'}_{\Delta} \right)\, ,\nonumber\\
F^{N,N'}_{9,9} &=& C(j_{\chi}) \frac{q^2}{16 m^2_N} F^{N,N'}_{\Sigma'}\, ,\nonumber\\
F^{N,N'}_{8,9} &=& C(j_{\chi}) \frac{q^2}{8 m^2_N} F^{N,N'}_{\Sigma', \Delta}\, ,
\eea
with $C(j_{\chi}) = 4 j_{\chi}(j_{\chi} + 1) /3$ being a prefactor that depends on the DM spin $j_{\chi}$ and has been normalized to $C(1/2) = 1$, and $q^2 = 2 m_T E_R$.

The astrophysics enters through the halo integral~\cite{Freese:2012xd}
\begin{eqnarray}
\eta(v_{\rm min},t) &=& \int_{v > v_{\rm min}}  \frac{f(\vec{v},t)}{v} d^3 v = \int_{v > v_{\rm min}} v f(\vec{v},t) dv d \Omega\, ,\nonumber\\
 h(v_{\rm min},t) &=& \int_{v > v_{\rm min}} v f(\vec{v},t) d^3 v = \int_{v > v_{\rm min}} v^3 f(\vec{v},t) dv d \Omega \, ,\nonumber\\
 \end{eqnarray}
where the minimum velocity required to produce a recoil energy $E_R$ in elastic scattering is 
\begin{equation}
v_{\rm min}(m_{\chi},E_R) = \sqrt{\frac{E_R m_T}{2\mu^2_{\chi T}} }\, ,
\end{equation}
with $\mu_{\chi T} = m_{\chi} m_T/(m_{\chi} + m_T)$ being the reduced mass.

\begin{figure}[!h]
\begin{center}
\includegraphics[width=\columnwidth]{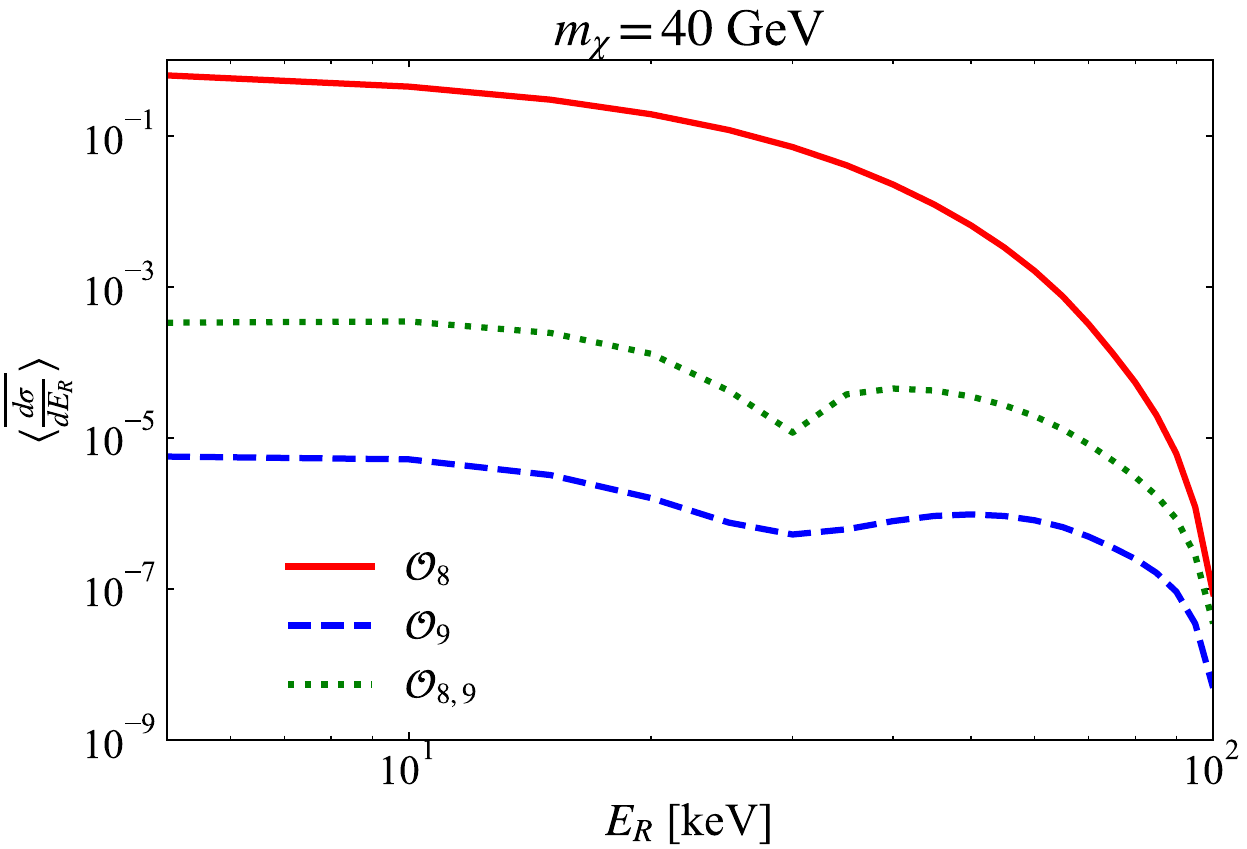}
\vspace*{-0.2cm}
\caption{The velocity and time averaged scattering rate, $\langle \overline{ \frac{d\sigma_{\chi T}}{d E_R} } \rangle$, up to a factor of $\frac{2 m_T}{4\pi} (c^p_8)^2$. Note that the green (dotted)  line shows the absolute value of the interference term, as $F^{(p,p)}_{\Sigma',\Delta}$ becomes negative when $E_R > 30$ keV.}
\label{fig:dsig-dER-mchi-40GeV}
\end{center}
\end{figure}

The velocity and time averaged scattering rate is a function of $m_{\chi}$ and $E_R$ 
\bea
\label{eq:sigma-v-t}
\langle \overline{\frac{d \sigma_{\chi T}}{d E_R}} \rangle &=& \frac{2 m_T}{4 \pi} (c^p_8)^2 \times \nonumber\\
&& \left\{  \frac{1}{4}\left( F_M^{(p,p)}(q^2) \cdot \Big[ \bar{h}(v_{\rm min}) - \frac{q^2}{4 m^2_T} \bar{\eta}(v_{\rm min}) \Big] \right. \right.\nonumber\\
&&\left. \left.\ \ \ \ \ \ \ \ \ \ \ \  + \frac{q^2}{m^2_N} F_{\Delta}^{(p,p)}(q^2) \cdot \bar{\eta}(v_{\rm min}) \right) \right.\nonumber\\
&& \left. \ + \frac{q^2}{16 m^2_N} F_{\Sigma'}^{(p,p)}(q^2) \cdot \bar{\eta}(v_{\rm min}) \right.\nonumber\\
&& \left. \ + \frac{q^2}{4 m^2_N} F_{\Sigma', \Delta}^{(p,p)}(q^2) \cdot \bar{\eta}(v_{\rm min}) \right\}\, ,
\eea
where $c^p_8 = c^p_9$ has been applied, and
\bea
\bar{\eta}(v_{\rm min}) &=& \int_0^1 \eta(v_{\rm min},t) dt\, ,\nonumber\\
\bar{h}(v_{\rm min}) &=& \int_0^1 h(v_{\rm min},t) dt\, .
\eea
We take $m_{\chi} = 40\ {\rm GeV}$ as an example, because this is the mass at which the direct detection experiments, such as PandaX~\cite{PandaX-4T:2021bab} and LZ~\cite{LZ:2024zvo}, set the strongest limits on both SI and SD cross sections. 
We compare the scattering rate (up to a factor of $\frac{2 m_T}{4\pi} (c^p_8)^2$) associated with ${\cal O}_8$, ${\cal O}_9$, and their interference in Fig.~\ref{fig:dsig-dER-mchi-40GeV}. From this figure we can see that Eq.~(\ref{eq:sigma-v-t}) is dominated by the effective operator ${\cal O}_8$ over a wide range of the recoil energy.
%

\bibliographystyle{apsrev4-2}
\bibliography{bibliography}

\end{document}